\documentclass[
aps,prb,
superscriptaddress, 
twocolumn,
longbibliography
]{revtex4-2}
\usepackage{appendix}
\usepackage{enumitem}
\usepackage{graphicx} 

\usepackage[usenames,dvipsnames]{xcolor}
\usepackage[
    bookmarksnumbered,
    pdfpagelabels=true,
    plainpages=false
    ]{hyperref}
\definecolor{cblue}{RGB}{55,126,184}
\hypersetup{
    colorlinks=true,
    linkcolor=cblue,
    citecolor=cblue,
    urlcolor=cblue
    }
\usepackage{array}
\usepackage{amsmath, amsfonts, amsthm}
\usepackage{amssymb,dsfont}
\usepackage{ulem}
\usepackage[capitalize]{cleveref}

\usepackage{cancel}
\usepackage{physics}
\usepackage{appendix}
\usepackage{xcolor}
\usepackage{tikz}
\usetikzlibrary{shapes,snakes}
\usetikzlibrary{positioning}
\usepackage[left=2cm,right=2cm,top=2cm,bottom=1.5cm]{geometry}

\makeatletter
\renewcommand*\env@matrix[1][\arraystretch]{%
  \edef\arraystretch{#1}%
  \hskip -\arraycolsep
  \let\@ifnextchar\new@ifnextchar
  \array{*\c@MaxMatrixCols c}}
\makeatother

\newcommand{\beq}{\begin{equation}}
\newcommand{\eeq}{\end{equation}}

\newcommand{\half}{\frac{1}{2}}
\newcommand{\threehalfs}{\frac{3}{2}}

\newcommand{\mH}{\mathcal{H}}
\newcommand{\tP}{\tilde{P}}
\newcommand{\tH}{\tilde{\mathcal{H}}}
\newcommand{\tsigma}{\tilde{\sigma}}

\makeatletter
\renewcommand{\p@subsection}{}
\renewcommand{\p@subsubsection}{}
\makeatother

\newcommand{\thistitle}{Generalized Wigner theorem for non-invertible symmetries}

\begin{document}

\title{\thistitle}

\author{Gerardo Ortiz}
\affiliation{Department of Physics, Indiana University, Bloomington, IN 47405, USA}
\affiliation{Institute for Advanced Study, Princeton, NJ 08540, USA}
\author{Chinmay Giridhar}
\affiliation{Department of Physics and Astronomy, Rice University, Houston, TX 77005, USA}
\author{Philipp Vojta} \affiliation{Department of Physics, Washington University in St. Louis, MO 63130 USA}
\author{Andriy H. Nevidomskyy}
\affiliation{Department of Physics and Astronomy, Rice University, Houston, TX 77005, USA}
\affiliation{Rice Center for Quantum Materials and Advanced Materials Institute, Rice University, Houston, TX 77005, USA}
\affiliation{Division of Condensed Matter Physics and Materials Science, Brookhaven National Laboratory, Upton, NY 11973-5000, USA}
\author{Zohar Nussinov}
\affiliation{Department of Physics, Washington University in St. Louis, MO 63130 USA}
\affiliation{Institut fur Physik, Technische Universität Chemnitz, 09107 Chemnitz, Germany
}
\affiliation{Department of Physics and Quantum Centre of Excellence for Diamond and Emergent Materials (QuCenDiEM),
Indian Institute of Technology Madras, Chennai 600036, India}

\date{\today}
\begin{abstract}

We establish the conditions under which a conservation law associated with a non-invertible operator may be realized as a symmetry in quantum physics.  As established by Wigner, all quantum symmetries must be represented by either unitary or antiunitary transformations. Relinquishing  
an implicit assumption of invertibility, we demonstrate that the fundamental invariance of quantum transition probabilities under the application of symmetries mandates that all non-invertible symmetries may only correspond to {\it projective} unitary or antiunitary transformations, i.e., {\it partial isometries}. This extends the notion of physical states beyond conventional rays in Hilbert space to 
equivalence classes 
in an {\it extended, gauged Hilbert space}, 
thereby broadening the traditional understanding of symmetry transformations in quantum theory. Our generalized theorem applies irrespective of the origin of the (non)invertible symmetry, holds in arbitrary spatial dimensions, and is independent of the Hamiltonian or action. We explore its physical consequences and, using simple model systems, illustrate how the distinction between invertible and non-invertible symmetries can sometimes be tied to the choice of boundary conditions.

\end{abstract}

\maketitle

\section{Introduction}
\label{section1}
Symmetry is a fundamental organizing principle underlying the laws of nature. At its core, symmetry refers to invariance, the idea that certain properties of a physical system remain unchanged under specific transformations, such as translations or rotations. This concept is not merely aesthetic; it has profound implications for the formulation of physical laws and the conservation principles they entail. For example, the invariance of physical laws under time translation leads to the conservation of energy, while rotational symmetry leads to conservation of angular momentum. As such, symmetry serves not only as a tool for understanding existing phenomena but also as a guiding principle in the pursuit of new physical theories.

In the context of quantum physics, one of the most profound consequences of symmetry is encapsulated in Wigner’s theorem \cite{Wigner_book}, including non-bijective (isometry) variants \cite{non-bijective}, which imposes strict constraints on the types of transformations that can be applied to quantum states. Specifically, the theorem asserts that {\it any symmetry transformation preserving transition probabilities between quantum states must be induced by either a unitary or an antiunitary isometry on the system’s Hilbert space}. This result not only ensures the mathematical consistency of quantum theory under symmetry operations \cite{Sakurai-Napolitano} but also reveals how deeply symmetries are woven into the fabric of the quantum realm, shaping everything from conservation laws to the subtleties of quantum measurement and the computation of event probabilities via the Born rule \cite{Asher-Peres}.

Recent developments in quantum many-body and field theories suggest the existence of symmetries that lie beyond the conventional framework of group theory, where each symmetry operation has an inverse. These {\it non-invertible symmetries}— also known as generalized or categorical symmetries— lack unique inverses and are better described using topological or higher-categorical structures \cite{MPO-lootens2023, lootens2024, Lootens2025}. They arise naturally in two-dimensional conformal field theories, topological quantum field theories, and certain lattice models, and are closely linked to phenomena such as duality transformations \cite{Cobanera2010,Cobanera2011}. 
This broader notion of symmetry hints at a deeper and more intricate structure underlying quantum systems which forms the nub of our work.

Guided by the analysis of one-dimensional
systems, Ref.~\cite{okada2024non} argued that non-invertible symmetries act locally on operators via completely positive maps \footnote{Completely positive linear maps cannot capture antiunitary symmetries such as time reversal.} realized by Stinespring dilations yet, as noted by its authors, without transition probability invariance (a ``weakened'' setting devoid of unitality or trace preservation).  Probabilities and probability invariance are, however, fundamental  cornerstones of quantum physics and its symmetries. To date, no study of non-invertible symmetries has investigated how transition probabilities may be conserved, nor identified the most general operator structure that a generalized symmetry—whether invertible or non-invertible—must possess, irrespective of spatial dimensionality or other system-specific details.  Indeed, at first glance, non-invertible symmetries appear to stand in tension with Wigner’s theorem. The aim of this paper is to resolve this apparent contradiction and propose a generalized extension of Wigner’s theorem that accommodates such symmetries in general systems.

\section{When Boundaries Matter}
\label{section2} 
We begin by illustrating the issue through well-known examples. A paradigmatic case is the transverse-field Ising chain (TFIC) \cite{NO2010,Fradkin-Susskind,Seo2025}, which encapsulates the essential features necessary to understand the apparent contradiction. At the outset, it is important to emphasize that, in the present case, the paradox arises from the specific boundary conditions imposed on the quantum systems  \cite{Cobanera2011}. 

Consider the TFIC with $L$ sites at the self-dual point associated with two   different boundary conditions \cite{Cobanera2011}: 
\begin{eqnarray}
    H_1&=&-\sum_{j=1}^{L-1} \sigma^z_j \sigma^z_{j+1} - \sum_{j=1}^{L-1} \sigma^x_j , \nonumber \\
    H_2&=&H_1- \sigma^x_L - \hat{\eta} \,\sigma^z_L\sigma^z_1 ,\label{eq:H2}
\end{eqnarray}
where $\boldsymbol{\sigma}_j=(\sigma^x_j,\sigma^y_j,\sigma^z_j)$ denotes Pauli matrices acting on site $j$, and
\begin{equation}
\hat{\eta} = \prod_{j=1}^L \sigma^x_j 
\end{equation}
is a $\mathbb{Z}_2$ symmetry common to both Hamiltonians. The first model ($H_1$) describes a TFIC with open boundary conditions while the closed TFIC ($H_2$) has a boundary term whose form depends on the eigenvalue of the symmetry operator $\hat{\eta}$.

What does it mean for $\hat{\eta}$ to be a symmetry? 
At the risk of sounding overly  pedantic, this means that (a) $\hat{\eta}$ is a conserved quantity, i.e., commutes with the Hamiltonian, \textit{and} (b) the action of $\hat{\eta}$ must preserve the transition probability between arbitrary quantum states. The latter requirement, which will become essential in what follows, is automatically satisfied if the operator is unitary or antiunitary -- the contents of Wigner's original theorem \cite{Wigner_book}. 

Do these Hamiltonians exhibit additional symmetries? As demonstrated in Ref. \cite{Cobanera2011}, both $H_1$ and $H_2$ possess a self-duality transformation (a linear automorphism of bond algebras \cite{Cobanera2010,Cobanera2011}) implemented by unitary maps whose action on an operator $\hat{O}$ is defined by $\Phi(\hat{O}) = U \hat{O}U^\dagger$, with $U$ unitary. Indeed, the unitary (and Hermitian) operator 
 \begin{eqnarray}
    {U}^{(1)} &=&  \left( \prod_{j=1}^{\lfloor \frac{L}{2}\rfloor} {\sf S}_{j,L-j+1} \right)\left( \prod_{j=1}^{L-1} {\sf C}^x_{j+1,j} \right){\sf H}^{\otimes L}  ,
    \label{U1}
 \end{eqnarray}
constructed entirely from elements of the Clifford group, including $\sf{Controlled}$-$\sigma^x$
\beq
\label{eq:CX}
\mathrm{C}^x_{i,j}
= e^{
i \frac{\pi}{4}\,
\left( \sigma^z_i \sigma^x_{j} - \sigma^z_i - \sigma^x_{j} +1 \right)}
,
\eeq
$\sf{Swap}$, ${\sf S}_{i,j}=\mathrm{C}^x_{i,j}\mathrm{C}^x_{j,i}\mathrm{C}^x_{i,j}=\frac{\mathds{1}+\boldsymbol{\sigma}_i\cdot \boldsymbol{\sigma}_j}{2}$ (which interchanges the spin at site $i$ with the one at site $j$), 
and Hadamard
\beq
\label{eq:Hadamard}
{\sf H}_{j} = i \, e^{-i \frac{\pi}{2\sqrt{2}} \left( \sigma^z_j + \sigma^x_j \right) },
\eeq
realizes the self-duality  map 
$\Phi_1(\sigma^x_j)=\sigma^z_{r_j}\sigma^z_{r_j+1}$, $\Phi_1(\sigma^z_{j}\sigma^z_{j+1})=\sigma^x_{r_j}$, $\Phi_1(\hat \eta)=\sigma^z_L$, with $r_j=L-j$ and $j=1,\cdots,L-1$, and commutes with $H_1$. Similarly, the unitary operator 
\begin{eqnarray}
    U^{(2)}&=&\Big ( \prod_{j=1}^{L-1} e^{-i\frac{\pi}{4}\sigma^x_j}
   e^{-i\frac{\pi}{4}\sigma^z_j\sigma^z_{j+1}}\Big ) e^{-i\frac{\pi}{4}\sigma^x_L} \nonumber \\
   &=&e^{-i\frac{\pi}{4}\sigma^x_1} e^{-i\frac{\pi}{4}\sigma^z_1\sigma^z_{2}}\cdots e^{-i\frac{\pi}{4}\sigma^x_L} , 
   \label{eq:U}
\end{eqnarray}
generates the map 
$\Phi_2(\sigma^x_j)=\sigma^z_{j}\sigma^z_{j+1}$, $\Phi_2(\sigma^z_{j}\sigma^z_{j+1})=\sigma^x_{j+1}$, with $j=1,\cdots,L-1$ and $\Phi_2(\sigma^x_L)= \hat \eta \,\sigma^z_L\sigma^z_1$, $\Phi_2(\hat \eta \,\sigma^z_L\sigma^z_1)=\sigma^x_1$, $\Phi_2(\hat \eta)=\hat \eta$, and 
commutes with the Hamiltonian $H_2$. 
Consequently, as argued in Refs. \cite{Cobanera2010, Cobanera2011}, these two unitary transformations qualify as respective {\it symmetries} of the system when the corresponding boundary conditions are implemented. We note that deriving explicit expressions for the unitary operators implementing the dualities is a notably challenging task. For example, even given $U^{(2)}$, deducing the form of $U^{(1)}$ is nontrivial and, to the best of our knowledge, has not appeared in the literature. The operator encodes the automorphism's entanglement characteristics and depends sensitively on the boundary conditions.

Consider now the same TFIC, but with explicit periodic or antiperiodic boundary condition:
\begin{equation}
H^\pm=H_1- \sigma^x_L \mp  \,\sigma^z_L\sigma^z_1.
\label{eq:H3}
\end{equation}

\noindent
The operators $U^{(1,2)}$ are no longer conserved quantities of $H^\pm$. 
In this case, no automorphism exists \cite{Cobanera2011};  instead one may attempt to define  {\it non-invertible operators}
\begin{eqnarray}
\label{eq:UP}
    D_\pm=U^{(2)} P_\pm ,
\end{eqnarray}
that, respectively, commute 
with the Hamiltonians $H^{\pm}$ of Eq.  \eqref{eq:H3}, i.e.,  $[H^{\pm},
D_\pm]=0$. 
Here, $P_\pm = (\mathds{1} \pm \hat{\eta})/2$ 
denote projection operators $(P^2_{\pm}= P_{\pm})$ onto the positive/negative symmetry eigenvalue $\eta=\pm 1$ sectors. Importantly, however, the transition probability between two arbitrary quantum states $| \alpha \rangle$, $|\beta \rangle$ in the Hilbert space ${\cal H}$ of dimension $2^L$ is no longer conserved after the action by $
D_{\pm}$. Indeed, explicitly implementing such a transformation,
\begin{eqnarray}
\label{eq:counterexample}
\big|\langle D_\pm\beta| D_\pm\alpha\rangle \big|^2=
\big| \langle \beta|P_\pm U^{(2) 
\dagger} U^{(2)} P_\pm| \alpha \rangle \big|^2 \nonumber \\ = 
\big|\langle \beta | P_\pm |\alpha \rangle \big|^2\ne \big|\langle \beta |\alpha \rangle \big|^2.
\end{eqnarray}
In other words, despite being a conserved quantity, $
D_\pm$ does \textit{not} qualify as a symmetry according to Wigner, since its probability altering action can be detected by measurements-- i.e., the application of $D_{\pm}$ does not lead to an invariance of the system. This example thus  underscores that 
even modest 
changes to the boundary
conditions may carry radical 
consequences for 
the nature (and existence) of the system automorphisms. 
As we will detail towards the end of this paper, \textit{bona fide} symmetry operators may, nonetheless, be explicitly written down for an amended (gauged) TFIC. 

\section{Generalized Wigner Theorem} 
\label{section3}

A key assumption in Wigner's theorem is that symmetry operators act on the same Hilbert space in which quantum states are defined; consequently, all such symmetries are inherently {\it invertible}. Since the preservation of transition probabilities is the central physical principle, we propose an extension of Wigner's theorem to include {\it non-invertible} (symmetry) operators. 

Our 
extension of Wigner's theorem 
to non-invertible symmetries asserts that:
{\it 
Any non-invertible symmetry 
transformation that preserves transition probabilities between quantum states of a  Hilbert space, can only be induced by the composition 
of either a unitary or an antiunitary operator and a projector onto a symmetry sector of an enlarged Hilbert space -- such that this projection operator acts as the identity on the original Hilbert space.}  Our proof applies to Hilbert spaces of finite or countably infinite (denumerably infinite) dimension.



\noindent 
{\bf Preliminaries}:

Consider a quantum system described by a Hermitian Hamiltonian $H$ 
acting on a Hilbert space $\cal H$ defined over the field of the complex numbers $\mathbb{C}$. The dimension of $\cal H$, denoted $\dim {\cal H}$, is assumed to be either finite or denumerably infinite. Any pair of arbitrary  normalized states $| \alpha \rangle$, $|\beta \rangle \in {\cal H}$ 
can be expanded  
 in 
 the orthonormal eigenbasis of $H$, i.e.,  $H | a_\mu \rangle = \epsilon_\mu | a_\mu \rangle$ ($\epsilon_\mu \in \mathbb{R}$ and $\langle a_\mu | a_{\nu}\rangle = \delta_{\mu\nu}$),
\begin{eqnarray}
\hspace*{-0.5cm}
 |\alpha\rangle=\sum_{\mu=1}^{\dim {\cal H}} {\sf a}_\mu | a_\mu\rangle  \ \mbox{ and }  |\beta\rangle=\sum_{\mu=1}^{\dim {\cal H}} {\sf b}_\mu | a_\mu\rangle ,  
 \end{eqnarray}
 with ${\sf a}_\mu\equiv \langle a_\mu| \alpha \rangle,{\sf b}_\mu\equiv \langle a_\mu| \beta \rangle \in \mathbb{C}$. The above system can, in general, be extended to a larger (finite or denumerably infinite of dimension $\dim \tH > \dim {\cal H}$) space $\tH={\cal H}_{\sf e}\otimes {\cal H}={\cal H}\oplus {\cal H}_\perp$, with normalized states $|\alpha_\perp\rangle  \in {\cal H}_\perp$, a subspace orthogonal to $\cal H$. The subsystem ${\cal H}_{\sf e}$ in the tensor product extension  is spanned by orthonormal states $\{ |g_\mu \rangle \}$. 
On this larger space $\tH$, the above Hamiltonian $H$ 
is replaced by a ``gauge-enlarged''  Hamiltonian $H_G$; 
the original Hamiltonian $H$ 
 may be understood as 
 $H = P_{\cal{H}} H_G P_{\cal{H}}$, where $P_{\cal{H}}$ is a projection operator onto  $\mathcal{H} \subseteq \tilde{\mathcal{H}}$. The lift to an enlarged Hilbert space considered here is not unique, yet it is not entirely arbitrary; the requirements of probability conservation and orthogonality to the remainder of the Hilbert space impose constraints on this extension.

We now return to our central focus: the investigation of non-invertible symmetries. In quantum physics, a physical property is invariant under unitary time evolution if and only if the corresponding operator $D$, defined on the physical Hilbert space $\mathcal{H}$,  commutes with the Hamiltonian 
\begin{eqnarray}
\label{commute''}
 [H, D] = 0,   
\end{eqnarray} 
where $D$ is represented by a bounded but not necessarily invertible operator.  
The corresponding operator ${\widehat D}$, defined on the extended space $\tH$, satisfies 
\begin{eqnarray}
 \label{DD'}
D=P_{\cal{H}}{\widehat D}P_{\cal{H}},    
\end{eqnarray}
meaning that the action of the $\mathcal{H}$-component of $\widehat{D}$ on states in $\mathcal{H}$ coincides exactly with the action of $D$ on those states. 
The action of $\widehat{D}$ on the pair of states $| \alpha \rangle$ and $| \beta \rangle \in {\cal H}$ is denoted by
\begin{eqnarray}
\label{D_action}
| \widetilde \alpha \rangle =  \widehat{D} |\alpha \rangle \ , \quad | \widetilde \beta \rangle =  \widehat{ D} |\beta \rangle .\end{eqnarray}
We stress that $\widehat D$ need not leave $\mathcal{H}$ invariant and thus $|\widetilde \alpha \rangle$ and $|\widetilde \beta \rangle$ may, generally, lie outside  ${\cal H}$. Recall that automorphisms may be linear or antilinear; antilinear maps can be written as the composition of a linear map with complex conjugation $K$.

For the conserved quantity $D$ to qualify as a symmetry, the associated transition probabilities -- defined via the inner product structure of the Hilbert space -- must remain invariant. In other words, it must hold that {\it for all} such arbitrary states $|\alpha \rangle, |\beta \rangle \in {\cal H}$, the conditions
\begin{eqnarray}
\label{transprob}
\hspace*{-0.6cm}  \big|\langle \widetilde \beta | \widetilde \alpha \rangle\big|^2 =  \big|\langle { \widehat{D}}\beta|\widehat{D}\alpha\rangle \big|^2=\big|\langle \beta | \widehat{D}^\dagger \widehat{D} |\alpha \rangle \big|^2  = \big|\langle \beta| \alpha \rangle \big|^2,
\end{eqnarray}
and, {\it for all} $|\alpha_\perp \rangle \in {\cal H}_\perp$,  
\begin{eqnarray}
\label{transprob'}
\hspace*{-0.6cm}  \big|\langle \widetilde \alpha_\perp | \widetilde \alpha \rangle\big|^2 =  \big|\langle { \widehat{D}}\alpha_\perp|\widehat{D}\alpha\rangle \big|^2=\big|\langle \alpha_\perp | \alpha \rangle \big|^2  = 0 ,
\end{eqnarray}
must be satisfied. The latter, nontrivial, condition emphasizes the key fact that upon a symmetry transformation, transitions between physical and unphysical  states are not allowed.




We next introduce our central result, a generalization of Wigner's theorem that is applicable to general symmetries (whether these symmetries happen to be invertible or non-invertible).

\noindent 
{\bf Theorem}: \\
A symmetry transformation will preserve the transition probabilities given in Eqs. \eqref{transprob} and \eqref{transprob'} if and only if it is of the form
\begin{equation}
\label{polar}
 \widehat{D} = \mathcal{U}\, \tP,   
\end{equation} 
where ${\cal U}$ is either unitary or antiunitary and $\tP$ is a positive semi-definite 
operator {\it that acts as the identity when restricted to} ${\cal H}$. Specifically, as we prove below, $\tilde{P}$ is  block diagonal with respect to the decomposition $\widetilde{\mathcal H} = \mathcal H \oplus \mathcal H_\perp$. It can be written as
\begin{eqnarray}
\label{Pblock}
\tP = P_{\cal H} + \tP_{{\cal H}_\perp},
\end{eqnarray}
where, $P_{\cal H}$ is the identity operator on ${\cal H}$ and 
$\tP_{{\cal H}_\perp}$ is a positive semi-definite operator acting solely within ${\cal H}_\perp$. 
Thus, whenever ${\cal H}_\perp$ is nontrivial, $\tP_{{\cal H}_\perp}$ has support only in ${\cal H}_\perp$.
Note that from the respective definitions of $\tP$ and the projection operator $P_{\cal{H}}$, it follows that $P_{\cal{H}} \tilde{P} = \tilde{P} P_{\cal{H}} = P_{\cal{H}}$.   \newline

\noindent {\bf Comment:}
Our theorem is {\it independent of the Hamiltonian} (Eq. (\ref{commute''})) {\it or action} defining the quantum theory. 
In both formulations, symmetries induce transformations that preserve transition probabilities. 
The theorem is further detached from other specific input details. In particular, the Hilbert space $\tH$ on which the theory is defined must include the physical space ${\cal H}$, viz., $\cal H \subseteq \tH$. 
The operators ${\cal U}$ and $\tP$  appearing in Eq. \eqref{polar} are defined on this general Hilbert space $\tH$. 
\newline


\noindent 
{\bf Proof}:

In what follows, as befits the ``if and only if'' nature of the Theorem, we explicitly break the proof into two parts. 
\begin{enumerate}[wide, labelwidth=!, labelindent=0pt]
\item
We first assume the form of Eq. \eqref{polar} for $\widehat{D}$ and aim to verify Eq. \eqref{transprob}. 
Indeed, 
for  $\tP$ that acts as the identity on arbitrary states $|\alpha \rangle, |\beta \rangle\in {\cal H}$ it then follows, from Eq. (\ref{D_action}), that the inner product 
\begin{eqnarray}
\label{eq:transition}
\langle \widetilde \beta | \widetilde \alpha \rangle &=&  
\langle \mathcal{U} \tP \, \beta  |\,
\mathcal{U} \tP \, \alpha \rangle = 
\langle \beta | \tP^\dagger \tP | \alpha \rangle = \langle \beta | P_{\mathcal{H}} | \alpha \rangle  \nonumber \\
&=&
\sum_{\mu=1}^{{\dim{\cal H}}}   {\sf b}^*_\mu {\sf a}^{\;}_\mu= 
\langle  \beta |  \alpha \rangle\  , 
\end{eqnarray} 
is preserved if ${\cal U}$ is unitary, where $^*$ denotes complex conjugation. Similarly, when ${\cal U}$ is antiunitary, we have
\begin{eqnarray}
\label{eq:antitransition}
\langle \widetilde \beta | \widetilde \alpha \rangle &=&  
\langle \mathcal{U} \tP \, \beta  |\,
\mathcal{U} \tP \, \alpha \rangle =
\langle \beta |\tP^\dagger \tP | \alpha \rangle^* = \langle \beta |  P_{\mathcal{H}}  | \alpha \rangle^* \nonumber \\
&=&
\sum_{\mu=1}^{{\dim{\cal H}}}   {\sf b}^{\;}_\mu {\sf a}^{*}_\mu= 
\langle  \alpha |  \beta \rangle, 
\end{eqnarray} 
i.e., probabilities are, in fact, conserved. Similar arguments readily establish that Eq. \eqref{transprob'} is satisfied.


\item
We next turn to the ``only if'' portion of our theorem and demonstrate that if probabilities remain invariant under the action of a 
transformation (a basic defining requirement of symmetries, following Wigner) then the corresponding operator {\it must be} of the form of Eq. (\ref{polar}). In what follows, we first discuss (i) the case of linear $\widehat{D}$ and then turn to (ii) antilinear $\widehat{D}$. \\
To illustrate this for {\it linear}  $\widehat{D}$, we recall that any 
linear operator (whether invertible or non-invertible) admits a (non-unique, left) polar decomposition (see, e.g.,  \cite{Wojcik_2017} and also Appendix \ref{section5}):
\begin{eqnarray}
\label{polar'}
{\widehat D}= \widehat{\cal U}\, {\widehat P}  \ ,
\end{eqnarray}
where $\widehat{\cal U}$ is unitary and ${\widehat P}$ represents a {\it positive semi-definite} (and thus Hermitian) operator. (When $\widehat{D}$ is non-invertible, $\widehat{P}$ has a non-trivial kernel.) 
Equation \eqref{polar'} implies that  $\widehat{D}^{\dagger} \widehat{D} =\widehat{P}^2$. Equation 
\eqref{transprob'} demands that $\forall \, |\alpha \rangle \in {\cal H}$ and, when a larger $\tH$ exists that includes ${\cal H}$ as a smaller subspace, $|\alpha_\perp \rangle \in {\cal H}_{\perp}$, the matrix elements $ \langle \alpha_\perp | \widehat{P}^2 | \alpha \rangle =0$, that is, $\widehat{P}^2$ may be expressed as a direct sum of an operator lying solely in ${\cal H}$ and (when $\tH$ is larger than ${\cal H}$) another operator that lies  exclusively in  ${\cal H}_{\perp}$, $\widehat{P}^{2} =  (\widehat{P}^{2})_{{\cal H}}+ (\widehat{P}^2)_{{\cal H}_{\perp}}$. Equation  (\ref{transprob}) reads   
\begin{eqnarray}
\label{P^2elements}
\big|\langle \beta | {\widehat{P}}^2|\alpha \rangle \big|^2=\big|\langle  \beta | \alpha \rangle\big|^2  \ .
\end{eqnarray}
Setting $| \beta \rangle = | \alpha \rangle$, we observe that $\langle \alpha | {\widehat{P}}^2 |\alpha \rangle =1$, $\forall \, |\alpha \rangle$, implying $\widehat{P}^2| \alpha \rangle = | \alpha \rangle$, given that $\langle \alpha_\perp | \widehat{P}^2 | \alpha \rangle =0$. In other words, within ${\cal H}$, the operator $\widehat{P}^2$ is the identity,  $P_{\cal H} \widehat{P}^2 P_{\cal H}= P_{\cal H}$. Thus the above direct sum for $\widehat{P}^2$ simplifies to  $\widehat{P}^{2} =  P_{{\cal H}}+ (\widehat{P}^2)_{{\cal H}_{\perp}}$. 
Given that $\widehat{P}$ is positive semi-definite, it must also enjoy the same block diagonal structure. Thus, $\widehat{P}$ must  be given by the operator $\tilde{P}$ of Eq. (\ref{Pblock}). 

We next address the case of 
{\it antilinear} non-invertible  symmetries.  
From the linear operator polar decomposition 
of Eq. (\ref{polar'}), 
all bounded antilinear symmetries  $\widehat{D}$ may  be written as 
\begin{eqnarray}
\label{antilinearD}
 \widehat{D} = \widehat{\cal U}\, {\widehat P} K. 
\end{eqnarray}
Replicating {\it mutatis mutandis} the above linear case proof 
yet now with the substitution of Eq. (\ref{antilinearD}) (instead of Eq. \eqref{polar'}) leads, once again, to the conclusion that 
$\widehat{P} =\tP$ must be 
a real 
operator that 
acts, for general states within ${\cal H}$, 
as the identity. 
Specifically, Eq. \eqref{transprob} implies that, within ${\cal H}$ the operator $ K \widehat{P}^2 K $ acts as the identity, $\langle \alpha |K \widehat{P}^2 K | \alpha\rangle=1$, for all $|\alpha\rangle \in {\cal H}$.
Therefore, in ${\cal H}$, the positive semi-definite 
$\widehat{P}$ itself must also act as the identity (
trivially commuting with  $K$). 
Hence, we may express antilinear symmetries $\widehat{D}$ in the form of Eq. (\ref{polar}) where ${\cal U}= \widehat{\cal{U}\,}K$ is now antiunitary and  $\widehat{P}=\tP$. 

The union of the two above ``if'' and ``only if'' portions of our proof concludes the demonstration of our theorem.  $\blacksquare$
\end{enumerate} 

\noindent {\bf Corollary 1:} 
For non-invertible symmetries $\widehat{D}$, the positive semi-definite operator $\tilde P_{\mathcal H_\perp}$  in Eq.~\eqref{Pblock} is non-invertible on $\mathcal H_\perp$.
This implies that {\it non-invertible symmetries can only exist if the physical Hilbert space is amended by an additional space ${\cal H _{\perp}}$}.
\newline
{\bf{Proof}}:  
If $\tilde P_{\mathcal H_\perp}$, viewed as an operator on $\mathcal H_\perp$, were invertible, with an inverse $(\tilde P_{\mathcal H_\perp})^{-1}$, then from Eqs. (\ref{polar},\ref{Pblock}) the symmetry $\widehat{D}$ would also be invertible, i.e.,
\begin{equation}
\widehat{D}^{-1} = \bigl(P_{\mathcal H} + (\tilde P_{\mathcal H_\perp})^{-1}\bigr) \, \mathcal U^\dagger.
\end{equation}
\noindent Thus, for the symmetry $\widehat{D}$ to be non-invertible, $\tP_{\cal H_{\perp}}$ cannot have an inverse.
 $\widehat{D}$ is non-invertible if and only if the positive semi-definite operator $\tilde{P}$ has a nontrivial kernel on the extended Hilbert space $\tH$, i.e., at least one of the eigenvalues of $\tP$  must vanish.  $\blacksquare$ \newline

Returning to the Hamiltonian framework, we obtain the following corollary: 

\noindent 
{\bf Corollary 2:} 
 The (anti)unitary operator $\mathcal{U}$ in Eq.~\eqref{polar}  
 must commute with the gauge enlarged Hamiltonian $H_G$ within $\mathcal{H}$, provided that  $[H_G,P_\mH]=0$.



\noindent
{\bf Proof:} 
From  Eq.~\eqref{commute''}, 
writing $\widehat{D}={\cal U} \tilde P$, and noting that $[H_{G}, P_{\cal{H}}]=0$, leads to the conclusion that $P_{\cal H}[ H_G, {\cal U} \tilde P] P_{\cal{H}}=0$. Substituting $P_{\cal{H}} \tilde{P} = \tilde{P} P_{\cal{H}} = P_{\cal{H}}$ and, once again, invoking the commutativity $[H_{G}, P_{\cal{H}}]=0$ immediately implies that $P_{\cal{H}}[H_{G}, {\cal{U}}] P_{\cal{H}}=0$. $\blacksquare$ \newline

The bond algebraic approach  \cite{Cobanera2010,Cobanera2011} is of great utility in constructing explicit realizations of the linear (or antilinear) map in Eq.~\eqref{polar} across different settings. In particular, defining the automorphism in the enlarged Hilbert space $\tilde{\mathcal{H}}$ guarantees that $[H_G, \mathcal{U}] = 0$ in $\tilde{\mathcal{H}}$.

\section{The Theorem at Work}

To close our circle of ideas and see how Eq. \eqref{polar} appears in a simple system, we return to the (anti)periodic TFIC 
which is now {\it minimally gauged} with a single $\mathbb{Z}_2$ gauge field $Z_{L+1}$ 
on the 
$(L,1)$ link (${\cal H}_{\sf e} = \mathbb{C}^2$),
\beq
 H_G = H_1 - \sigma^x_L - \sigma^z_L Z_{L+1} \sigma^z_{1} . 
 \label{eq:H_min_gauged}
\eeq 
This enlarges the Hilbert space $\cal H$ by a factor of two, yielding $\tilde{\mathcal{H}} =\mathbb{C}^2  \otimes \mathcal{H}$, with dimension $\dim \tilde{\mathcal{H}} = 2^{L+1}$. Projecting onto the gauge sector $Z_{L+1} = 1 (-1)$ gives us $H^{+(-)}$ which acts on ${\cal H} \, ({\cal H}_\perp)$. These TFICs are known to possess the non-invertible conserved quantity $D_\pm$ (Eq. \eqref{eq:UP}), implementing the self-duality. The motivation behind considering the minimally gauged model $H_G$ is the existence of a bond-algebraic automorphism~\cite{Cobanera2010,Cobanera2011} $\Phi_G$ implementing the self-duality on the enlarged Hilbert space, 
such that $[H_G,{\cal U}]=0$ with unitary $\cal U$. Precisely, it acts as:
$\Phi_G(\sigma^z_j \sigma^z_{j+1}) = \sigma^x_{j+1}$, $\Phi_G(\sigma^x_{j}) = \sigma^z_{j} \sigma^z_{j+1} , \,  j=1,\ldots, L-1$, and the boundary bonds $\Phi_G(\sigma^z_L Z_{L+1} \sigma^z_1) = \sigma^x_1$, 
and $\Phi_G(\sigma^x_L) = \sigma^z_L Z_{L+1} \sigma^z_1$. Following Ref.~\cite{seiberg-LSM}, that unitary operator is given by 
 \beq
    \mathcal{U} = {\sf C}^x_{1,L+1} \left( \prod_{j=1}^{L} {\sf H}_j {\sf C}^z_{j+1,j} \right) {\sf H}_{L+1}.
    \label{eq:U_min_gauged}
 \eeq

To explicitly connect this unitary operator 
to $U^{(2)}$ in Eq. \eqref{eq:U}, 
we express the various factors in Eq. \eqref{eq:U_min_gauged} in terms of the local Pauli operators. From Eq.~\eqref{eq:CX} we have 
\[
{\sf C}^x_{1, L+1}
= e^{
i \frac{\pi}{4}\,
\left( \sigma^z_1 X_{L+1} - \sigma^z_1 - X_{L+1} +1 \right)}
,
\]
the local {\sf Controlled}-$\sigma^z$ Clifford operations are given by 
\beq
{\sf C}^z_{j+1,j} = e^{-i \frac{\pi}{4} \left( \sigma^z_j \sigma^z_{j+1} - \sigma^z_j - \sigma^z_{j+1} +1 \right) },
\eeq
while the Hadamard operator is defined in Eq.~\eqref{eq:Hadamard},
with the understanding that when acting on the last link ($j=L+1$) one should interpret $\sigma^{x}$ ($\sigma^{z}$) operators as the gauge fields $X_{L+1}$ ($Z_{L+1}$) respectively.

The non-invertibility of the operator $\widehat{D} = \mathcal{U} \tP$ stems from the 
projection operator $\tP$, which is defined on the extended Hilbert space by its action on the gauge degree of freedom: $\tP_{\pm} = (\mathds{1} \pm Z_{L+1})/2$, while acting trivially on the original Hilbert space, as required by the theorem. Both the physical space ${\cal H}$ and its orthogonal complement ${\cal H}_\perp$ have dimension $2^{L}$.  In Eq.~(\ref{Pblock}), $P_{\cal H}$ is the identity operator on ${\cal H}$ while, restricted to ${\cal H}_{\perp}$, the operator $\tilde{P}_{\cal H_{\perp}}$ is the (trivially non-invertible) $2^L \times 2^L$ dimensional zero operator.

We note that if one attempted instead to 
define the operator $\widehat D = \mathcal{U} P$ yet with a projector $P$ that {\it acts non-trivially} on the original Hilbert space 
$\mathcal{H}$, 
such a non-invertible operator will, generally, {\it not} preserve transition probabilities. An explicit example is that of Eq.~\eqref{eq:UP} for the TFIC. 
That is why one must go to the trouble of defining a gauged Hamiltonian in Eq.~\eqref{eq:H_min_gauged} and the associated enlarged Hilbert space $\tH$.

In Appendix \ref{section6}, we demonstrate that an alternative global gauging of the original Hamiltonian $H^\pm$ is spectrally equivalent —up to uniform degeneracies— to our minimally gauged Hamiltonian $H_G$. This equivalence is established via a dual mapping between the two Hamiltonians, indicating that all information about the isometry is encoded in $H_G$. In this way we can freely enlarge the kernel of $\tilde{P}$  while  ${\mathcal{U}}$ remains fixed.


\section{Conclusions}
\label{section4}

In earlier investigations on non-invertible symmetries, a defining focus has been the need to satisfy 
$[H,{D}] = 0$ to ensure that ${D}$ is a conserved quantity. Building on the central idea underlying Wigner’s original theorem, {\it quantum measurement} imposes the fundamental physical requirement that any symmetry, including a non-invertible one, must leave all transition probabilities invariant.
This led us to the rigorous conclusion that all non-invertible symmetries that preserve transition probabilities in Eq.~(\ref{transprob}) must take the form of a unitary or antiunitary operator compounded by a positive semi-definite  operator $ \tP$ that acts as the identity on physical states, as expressed in Eq. (\ref{polar}). In other words, they are  {\it partial isometries} that  correspond to the gauge-reducing/enlarging dualities  introduced in Ref. \cite{Cobanera2011}. Crucially, we show that $\tP$ cannot be  a nontrivial projector on the original Hilbert space ${\cal H}$ (i.e., projecting from the Hilbert space ${\cal H}$ to a smaller subspace ${\cal H}_{\sf sub} \subset {\cal H}$). If $\tP$ were  such a projection operator to a smaller subspace, then transition probabilities would not be preserved, as per Eq.~\eqref{eq:counterexample}; the transformation of Eq.~\eqref{eq:UP} 
for the periodic TFIC exemplifies the case in point.  As we further expounded on the TFIC example, the non-invertibility of the symmetry is sensitive to (and may be triggered by) the choice of the boundary conditions. In particular, a non-invertible symmetry can be rendered to be of a familiar unitary form by a particular choice of boundary conditions, as exemplified by $H_1$ and $H_2$ in Eq.~\eqref{eq:H2}. 

Elsewhere~\cite{Chinmay-XM}, building on the operator based  bond-algebraic technique (distinct from Kramers-Wannier-type derivations of dualities based on loop-counting series expansion), we illustrate this to be the case for other systems. Our theorem, however, is independent of any assumptions about the source of the non-invertibility, irrespective of whether it originates from boundary conditions. The strength of the bond-algebraic duality technique \cite{Cobanera2010, Cobanera2011} lies in its ability to explicitly identify and construct  automorphisms $\cal U$. 

Following Wigner’s foundational insight, we emphasize that physical states are not represented by individual vectors in a Hilbert space, but rather by equivalence classes of vectors. In Wigner's case, vectors differing by an overall phase are equivalent -- corresponding to the same \textit{ray} -- reflecting the fact that global phase factors are physically unobservable. In the context of non-invertible symmetries, this observation necessitates extending the notion of rays to accommodate gauging. Thus, a proper treatment of generalized symmetries requires a corresponding generalization of the concept of physical states beyond simple Hilbert space vectors, incorporating the additional structure introduced by gauge redundancy: 
$$\boxed{\mbox{\! \sf Physical states}\leftrightarrow 
{\sf Equivalence}\, {\sf classes}\, |\Psi\rangle \simeq e^{i \gamma} |g\rangle \!\otimes \!|\alpha\rangle}$$
where $|g\rangle$ denotes the specific 
sector in the (enlarged) gauged Hilbert space.
%
This requires a mathematical structure akin to the concept of a vector bundle, and is reminiscent of the fiber bundles (and related field-bundles) used in describing gauge field theories~(see Ref.~\cite{schreiber-bundles} for a pedagogical exposition).  Different gauge choices correspond to the sections of the field bundle.  

Our analysis reveals a fundamentally new perspective. 
At its core, the theorem concerns quantum measurements: it shows that uncovering {\it any} non-invertible symmetries requires an {\it active} observer, ``Wigner's agent,'' whose role is to enlarge the Hilbert space and perform a projective measurement on it, explicitly incorporating  active observation into the quantum description.
By contrast, Wigner’s original theorem applies to a {\it passive} observer and therefore captures only invertible symmetries. This distinction offers new physical insight into how symmetries arise in quantum systems and highlights the essential role of the observer in revealing them through measurement.


Our theorem 
leads to practical consequences in several arenas. 
In quantum information, particularly 
circuit constructions of non-invertible symmetries, 
our theorem asserts that non-unitary operations, such as projective measurements, cannot act on the qubit representation of the physical system. 
Instead, these measurements must be performed on suitably chosen auxiliary qubits (ancillae). The TFIC provides a concrete realization 
of this 
maxim. 
Naively inserting the projector $P_{+}$ into the duality operator (Eq. \eqref{eq:UP}) would incorrectly alter the transition probabilities (Eq. \eqref{eq:counterexample}). The correct implementation necessarily enlarges the state space with ancillae, as shown in Eq. \eqref{eq:H_min_gauged}. 

Our results also inform 
future applications in quantum simulation and laboratory probes of non-invertible symmetries. By analogy with the classic neutron-interferometry experiment confirming that the spin-1/2 wavefunction realizes a projective representation of the rotation group SO(3) \cite{Rauch75} -- where Wigner’s theorem fixed the rotation operator to be strictly unitary --  an experiment aiming to measure the action of any non-invertible symmetry must implement the duality operator in the constrained form $\widehat{D} = \mathcal{U}\tilde{P}$ (Eq. \eqref{polar}). This constraint should serve as a guiding principle in the design of such experimental setups.

\begin{acknowledgments} 
We thank Thomas Bartsch for finding our generalized Wigner theorem interesting  and asking for further details just after our work first appeared on the arXiv. 
 G.O. gratefully acknowledges support from the Institute for Advanced Study. Z.N. is grateful for support from TU Chemnitz and the IIT Madras. Part of this work was performed at the Aspen Center for Physics, which is supported by National Science Foundation grant PHY-2210452. C.G. and A.H.N. were supported by the Department of Energy under the Basic Energy Sciences award no. DE-SC0025047. 
\end{acknowledgments}

\appendix

\section*{APPENDICES}

Here, we provide a simple proof of the polar decomposition for non-invertible operators, which plays a central role in the formulation of the main theorem. Furthermore, we offer a conceptual justification for the minimal gauging strategy employed in the main text, drawing on the structure and properties of duality maps to support our approach.

\section{Polar decomposition of operators}
\label{section5}

The polar decomposition theorem (Eq. (\ref{polar'})) is well known and may be readily demonstrated by various inter-related approaches. 
This decomposition is most often discussed and employed for invertible operators. Since the use of this general theorem (applicable to all operators whether they happen to be invertible or not) is germane to our central result, for completeness and clarity (using compact Dirac notation), we pedagogically provide a proof that reviews, in passing, the Singular Value Decomposition for {\it{arbitrary}} operators $A$. 

Given such general operators, the two products $(A^{\dagger} A)$ and $(AA^{\dagger})$ are Hermitian. Thus, these two operators  trivially enjoy a spectral decomposition into complete basis spanned by sets (respectively, $\{|v_{\mu} \rangle \}$ and $\{|w_{\mu} \rangle\}$) of orthonormal eigenvectors, 
\begin{eqnarray}
\label{ortho}
  \langle v_{\mu}| v_{\nu} \rangle = \langle w_{\mu} | w_{\nu} \rangle = \delta_{\mu\nu},
\end{eqnarray} 
that is,
\begin{eqnarray}
\label{AA}
A^{\dagger} A|v_{\mu} \rangle &=& \lambda_{\mu} | v_{\mu} \rangle , \nonumber \\
AA^{\dagger}|w_{\mu} \rangle &=& \lambda_{\mu} |w_{\mu} \rangle , \end{eqnarray}
with $\lambda_\mu \in \mathbb{R}$. 
In the outer product basis formed by these complete orthonormal states, the general operator $A$ may be consistently expressed as 
\begin{eqnarray}
\label{Asum}
A = \sum_{\mu} \sigma_{\mu}  |w_{\mu} \rangle \langle v_{\mu} |,  
\end{eqnarray}
with $\lambda_{\mu} = | \sigma_{\mu}|^2$.
Equation (\ref{Asum}) implies, for these orthonormal bases vectors, that $A|v_{\mu} \rangle = \sigma_{\mu} |w_{\mu} \rangle$ -- a generalization of the eigenvector equation 
to also include operators $A$ that do not necessarily have eigenvectors. By judiciously multiplying individual eigenvectors $|v_{\mu} \rangle$ and $|w_{\mu} \rangle$ by appropriate phase factors in Eqs. (\ref{AA}) (doing so leaves Eqs. (\ref{AA}) invariant), we can set all ``singular values'' $\{\sigma_{\mu}\}$ to be non-negative, $\sigma_{\mu} \ge 0$. Defining operators $V$ (and $W$) whose columns are given by $\{|v_{\mu}\rangle\}$ (respectively, $\{|w_{\mu} \rangle\}$), and a diagonal operator $\Sigma$ whose eigenvalues are the non-negative $\{\sigma_{\mu} \}$, we may recast the Singular Value Decomposition of Eq. (\ref{Asum}) into the more common (non-Dirac notation) form appearing in linear algebra textbooks, viz., 
\begin{eqnarray}
\label{SVD}
A = W \Sigma V^{\dagger} . 
\end{eqnarray}
Here, given the orthonormality of their column vectors (Eq. (\ref{ortho})), the operators $W$ and $V$ are unitary. 

We underscore that no assumptions have been made in deriving the standard Singular Value Decomposition form of Eq. (\ref{SVD}). As such and as is well appreciated, the Singular Value Decomposition indeed applies to {\it all operators} $A$ including those operators having a rectangular matrix representation (in which case, $\Sigma$ is represented by a rectangular matrix and $W$ and $V$, generally,  by matrices of different dimension since in Eqs. (\ref{AA}),  $(AA^{\dagger})$ and $(A^{\dagger} A)$ are Hermitian operators of different dimension). Our focus is, however, on physically pertinent square matrix representations $A$ in which $W, \Sigma,$ and $V$ become square matrices of equal dimension (with the number of columns/rows equal to the dimension of the Hilbert space $\tilde{\cal H}$ discussed in the main text). Given that $W$ and $V$ are unitary (and thus have trivial inverses $W^{-1}=W^{\dagger}$ and $V^{-1}=V^{\dagger}$), the Singular Value Decomposition of Eq. (\ref{SVD}) implies that the operator $A$ does not have an inverse if and only if at least one of the singular values vanishes, $\sigma_{\mu}=0$. Equivalently, $A$ is non-invertible if and only if the singular matrix $\Sigma$ has 
at least one null vector -- that is, $\Sigma$ has a nontrivial kernel. Indeed, if all singular values are positive $\sigma_{\mu} >0$ then it follows from Eq. (\ref{SVD}) that the operator $A$ is invertible. 

As we emphasized earlier, the polar decomposition theorem (employed in Eq. (\ref{polar'}) of the main text) applies to all operators (whether invertible or non-invertible). 
The derivation of this theorem from the Singular Value Decomposition of Eqs. \eqref{Asum} and \eqref{SVD} is immediate. To see this, we may simply set  
\begin{eqnarray}
\widehat{\cal U} \equiv W V^\dagger, \quad \widehat{P} \equiv V \Sigma V^\dagger.
\label{PHermitian}
\end{eqnarray}
The unitarity of the so-defined operator $\widehat{\cal U}$ is evident given that both $W$ and $V$ are unitary. Furthermore, since $\Sigma$ is a diagonal real operator with non-negative entries and $V$ is unitary, it readily follows that $\widehat{P}$ is a Hermitian operator. Returning to our earlier review of the Singular Value Decomposition, if $A$ is non-invertible then $\widehat{P}$ has a null space (a nontrivial kernel) and is a positive semi-definite operator. Conversely, if $A$ is invertible then $\widehat{P}$ is a positive definite operator.    

Thus, putting all of the pieces together, it is seen from the  Singular Value Decomposition of Eq. (\ref{SVD}) -- a decomposition that is applicable to all operators -- that
any operator $A$ can be written as \begin{eqnarray}
A = \widehat{\cal U} {\widehat{P}},
\end{eqnarray}
with $\widehat{\cal U}$ denoting a unitary operator and ${\widehat{P}}$ a positive semi-definite one.
In the main text, we focused on  
operators $A= {\widehat D}$ (Eq. (\ref{polar'})) with 
the Hermitian operators of ${\widehat D}^\dagger {\widehat D}$
whose elements determine the transition probabilities (Eq. (\ref{transprob})) corresponding to the general Hermitian operators of Eq. (\ref{AA}). In such instances, $\widehat{P}$ is a positive semi-definite operator, and the discussion after Eq. (\ref{polar'}) follows.

\section{From full gauging to \\ minimal gauging via duality}
\label{section6}


In the main text, we present an example demonstrating the application of the generalized Wigner theorem to the so-called minimally-gauged TFIC, Eq.~\eqref{eq:H_min_gauged}. Here, we show how this model can be reached starting from a fully-gauged TFIC model, where $\mathbb{Z}_2$ gauge degrees of freedom $\tilde{X}_{j+\half}$ are placed on every link \mbox{$(j,j+1)$} of the chain, 
\beq
\tilde H_G = -\sum_{j=1}^{L-1} (\sigma^z_j   \tilde{X}_{j+\half}\sigma^z_{j+1} + \sigma^x_j) - \sigma_L^x - \sigma^z_L \tilde{X}_{\half} \sigma^z_1,
\eeq
subject to the Gauss law on every site $G_j = \tilde{Z}_{j-\half} \sigma^x_j \tilde{Z}_{j+\half} = 1$.
At first glance, it might appear that we have enlarged the Hilbert space to having dimension $2^{2L}$. However, the Gauss laws enforce the gauge redundancy, resulting in a smaller physical Hilbert space.
To see this explicitly, we introduce the dual variables containing a string of $\tilde{X}$'s in their definition \cite{seiberg-LSM}
\beq
\tsigma^z_j = \tilde{X}_\half \tilde{X}_\threehalfs\ldots \tilde{X}_{j-\half}\sigma_j^z;\quad  \tsigma^x_j = \sigma_j^x. 
\eeq
Note that such variables have a long history, closely related to the disorder variables in the context of TFIC~\cite{Fradkin-Susskind,Cobanera2011}. 
%
In terms of these new matter fields (replacing $\tsigma \to \sigma$ for clarity), the gauged Hamiltonian becomes 
\begin{eqnarray}
H_G &=& -\sum_{j=1}^{L-1} (\sigma^z_j   \sigma^z_{j+1} + \sigma^x_j) - \sigma^x_L - \sigma^z_L  Z_{L+1} \sigma^z_1\nonumber \\
&=& H_1 - \sigma^x_L - \sigma^z_L Z_{L+1} \sigma^z_1,
\label{eq:H_G}
\end{eqnarray}
where 
\beq
Z_{L+1}  = \prod_{j=1}^L \tilde{X}_{j-\half}. 
\label{eq:L+1}
\eeq
Crucially, the only remnant of the gauge degrees of freedom enters via the last $(L,1)$ link. This is exactly the minimally-gauged model in Eq.~\eqref{eq:H_min_gauged}. 
It follows that it is therefore sufficient to enlarge the Hilbert space by one ancilla qubit: $\tilde{\mH} = \mathbb{C}^2 \otimes \mH$, with dimension $\mathrm{dim} \, \tilde{\cal H}=2^{L+1}$, thus justifying the minimal gauging in Eq.~\eqref{eq:H_min_gauged}.

Proceeding as in the main text, we now define the projector operator $\tP_\pm = (\mathds{1} \pm Z_{L+1})/2$ which projects onto a definite ancilla eigenvalue $\tilde{\eta} = \pm 1$, while acting trivially on the original Hilbert space:  $\tP|_\mH = P_\mH$. 
As shown in the main text, the operator $\widehat{D}_\pm = {\cal U} \tP_\pm$, with $\cal U$ defined in Eq.~\eqref{eq:U_min_gauged} commutes with the Hamiltonian $H_G$ above, thus implementing the non-invertible symmetry as required by the generalized Wigner theorem.

Several remarks are in order: (i) $\tilde{\mH}$ contains two copies of the original Hilbert space, distinguished by the eigenvalue $\tilde{\eta}=\pm 1$ of the ancilla: $\tH = \mH \oplus \mH$, and (ii) the Hamiltonian in the ``twisted'' sector, corresponding to $\tilde{\eta}=-1$, contains an $\eta$-defect on the $(L,1)$ link, relative to the sign on the rest of the bonds in $H_1$. Adopting the language of Ref.~\cite{seiberg-LSM}, the Hamiltonian $H_G$ in the enlarged Hilbert space $\tH$ can thus be interpreted as containing an $1\oplus \eta$ defect -- this observation is central to the interpretation of $1 \oplus \eta = D\otimes D$ as the result of fusing two $D$-defects in Ref.~\cite{seiberg-LSM}.

Finally, we point out that (iii) the Hamiltonian $H_G$ in Eq.~\eqref{eq:H_G} bears, modulo a replacement of the global symmetry operator \mbox{$\hat{\eta} = \prod_{j=1}^L \sigma^x_j$} by the emergent
\textit{dual symmetry} operator $\widehat{\tilde{\eta}} = \prod_{j=1}^L \tilde{X}_{j-\half} \equiv Z_{L+1}$ in Eq.~\eqref{eq:L+1}, an uncanny resemblance to $H_2$ in Eq.~\eqref{eq:H2}. This echoes the observation, made in Ref.~\cite{seiberg-LSM}, that the  duality implemented by the operator $\widehat{D}_\pm = U\tP_{\pm}$ interchanges $\eta \longleftrightarrow \tilde{\eta}$. Physically, the eigenvalue $\eta$ marks the (linear superposition of) the symmetry-broken ground states in the ordered phase, while $\tilde{\eta}$ corresponds to the so-called disorder variable. The $D$-duality thus interchanges the order and disorder variables of the TFIC.

\sloppy
\renewcommand{\emph}{\textit}
\let\originalunderline\underline

\renewcommand{\underline}{\uline}
\bibliography{refs}
\let\underline\originalunderline

\end{document}